\documentclass[prl, twocolumn, showpacs]{revtex4}
\usepackage{amsmath}
\usepackage{amsfonts}
\begin{document}
\title{Positronium collapse and the maximum magnetic field in pure
QED}
\author{Anatoly E. Shabad }
\affiliation{P.N. Lebedev Physics Institute, Moscow 117924,
Russia}
\author{Vladimir V. Usov}
\affiliation{Center for Astrophysics, Weizmann Institute of
Science, Rehovot 76100, Israel}
\begin{abstract}
A maximum value for the magnetic field is determined, which
provides the full compensation of the positronium rest mass by the
binding energy in the maximum symmetry state and disappearance of
the energy gap separating the electron-positron system from the
vacuum. The compensation becomes possible owing to the falling to
the center phenomenon. The maximum magnetic field may be related
to the vacuum and describe its structure.
\end{abstract}
\pacs{03.65.Ge, 03.65.Pm, 12.20.Ds, 98.80.Cq}
\maketitle
A common feature of compact astronomical objects: white dwarfs,
neutron stars, and accretion disks around black holes, is a very
strong magnetic field. It is believed that neutron stars possess
the strongest observed magnetic fields. The field strength is in
the range from $\sim 10^8$ G to $\sim 10^{14}$ G for radio pulsars
identified with rotation-powered neutron stars \cite{M03}, and may
be as high as $\sim 10^{15}$ G for soft gamma-ray repeaters
\cite{TD95}, or even higher ($\sim 10^{16}- 10^{17}$ G) for the
sources of cosmological gamma-ray bursts \cite{U92}. Much more
intense magnetic fields have been conjectured to be involved in 
several astrophysical phenomena. For instance, superconductive
cosmic strings, if they exist, may have magnetic fields up to
$\sim 10^{47}-10^{48}$ G in their vicinities \cite{W95}. Magnetic
fields of $\sim 10^{47}$ G may be also produced in our Universe at
the beginning of the inflation \cite{KK89}. The fundamental
physical problems are how large the field strength can be in
nature, and how the properties of the vacuum change when magnetic
fields approach the extremity. It is accepted that magnetic fields
are stable in pure quantum electrodynamics (QED), and another
interaction (weak or strong) or magnetic monopoles have to be
involved to make the magnetized vacuum unstable \cite{NO78}.

In this Letter, working solely in QED, we find that there exists
a maximum value of the magnetic field that delimits the range of
its values admitted without
revising QED. Its value is many orders of magnitude less than
$B=B_0\exp (3\pi/\alpha)$ (here $\alpha=e^2/4\pi\simeq 1/137,$
$B_0=m^2/e=4.4\times 10^{13}$ Gauss, $m$ is the electron mass,
$\hbar=c=1$ throughout), the value that restricts the range of
validity of QED due to the lack of asymptotic freedom
\cite{ritus1}. The maximum magnetic field causes the shrinking of
the energy gap between an electron and positron.

We exploit the unboundedness from below of the energy spectrum for
sufficiently singular attractive potentials, known as the "falling
to the center" \cite{QM}. For $e^+e^-$ system  
in a magnetic field $B$ this phenomenon is caused by the
ultraviolet singularity of the photon propagator and occurs in the
limit $B\rightarrow\infty$. It takes place independently of
whether nonrelativistic or relativistic description is used, and
is associated with dimensional reduction due to the charged
particles being restricted to the lowest Landau levels (see the
pioneering work of Loudon \cite{loudon} and other works out of
which  most important for our present theme are \cite{loudon} -
\cite{gusynin}). Using the Bethe-Salpeter (BS) equation for the
electron-positron system, 
with the relative motion of the two particles treated in a
strictly relativistic way, we show that at a maximum value
of the magnetic field, which may be of an astrophysical
significance, the rest energy of the system is compensated
for by the mass defect,
i.e., the system is not separated from the vacuum by an energy
gap.
We refer to this situation as a collapse of positronium.

In processing the formalism and interpreting the results,
especially while discussing the vacuum structure, we use the
theory of the falling to the center developed in \cite{shabad}
that implies deviations from the standard quantum theory
manifesting themselves when extremely large electric fields near
the     singularity     become     important     \cite{footnote1}.

We proceed from the (3+1)-dimensional BS equation in an
approximation, which is the ladder approximation once the photon
propagator (in the coordinate space) is taken in the Feynman
gauge: $D_{ij}(x)=g_{ij}D(x^2)$, $x^2\equiv x_0^2-{\bf x}^2$,
$g_{ij}$ is the metrics, $g_{ii}=(1,-1-1-1)$. In an asymptotically
strong magnetic field this equation may be written in the
following (1+1)-dimensional form (see our detailed paper
\cite{SU05} for the derivation), covariant under the Lorentz
transformations along the axis 3:
\begin{eqnarray}\label{closed1}({{\rm i}\overrightarrow{\hat{
\partial_\|}}- \frac{\hat{P_\parallel}}2-m})
\Theta(t,z)(-{\rm i}\overleftarrow{\hat{
\partial_\parallel}}- \frac{\hat{P_\parallel}}2-m)\quad\quad\quad
\nonumber\\ = {\rm i}
8\pi\alpha~\sum_{i=0,3}D\left(t^2-z^2-\frac{P_\perp^2}{(eB)^2}
\right)g_{ii}\gamma_i
\Theta(t,z)\gamma_i,
\end{eqnarray}
where $\Theta (t,z)$ is the 4$\times 4$ (Ritus transform of) BS
amplitude, $t=x^{\rm e}_0- x^{\rm p}_0$ and $z=x^{\rm e}_3-x^{\rm
p}_3$ are the differences of the coordinates of the electron (e)
and positron (p) along the time $x_0$ and along the magnetic field
${\bf B}=(0,0,B_3=B)$, respectively. $P_\|$ and $P_\perp$ are
projections of the total (generalized) momentum of the positronium
onto the (0,3)- subspace and the (1,2)-subspace, respectively.
Only two Dirac gamma-matrices, $\gamma_{0,3},$ are involved,
$\hat{
\partial_\|}=\gamma_0\partial/\partial_t+
\gamma_3\partial/\partial_z$,
$\hat{P_\|}=P_0\gamma_0-P_3\gamma_3$.

Equation (\ref{closed1}) is valid in the coordinate domain, where
the argument of the function $D$ is greater than the electron
Larmour radius squared $(L_B)^2=(eB)^{-1}$. When $B=\infty$, the
domain of validity covers the whole exterior of the light cone
$z^2-t^2\geq 0$. The Lorentz-invariant expansion  of $\Theta$ over
matrix basis contains four independent scalar components, whose
number diminishes to three if $P_\|=0$. The argument of the
original photon propagator $(x^{\rm e}-x^{\rm p})^2$ has proved to
be replaced in (\ref{closed1}) by
$t^2-z^2-(\widetilde{x}_\perp^{\rm e}-\widetilde{x}_\perp^{\rm
p})^2 =t^2-z^2-P_\perp^2/(eB)^2$, where $\widetilde{x}_\perp^{\rm
e,p}$ are the center of orbit coordinates of the two particles in
the transverse plane. Now that after the dimensional reduction
this subspace no longer exists these substitute for the transverse
particle coordinates themselves: $\widetilde{x}_\perp^{\rm
e,p}$are not coordinates, but quantum numbers of the transverse
momenta. The mechanism of replacement of
a coordinate by a quantum number is the same as 
in \cite{ShUs}.

In deriving equation (\ref{closed1}) the expansion over the
complete set of Ritus matrix eigenfunctions \cite{ritus} was used
in \cite{SU05} that accumulate the dependence on the transverse
spatial and spinorial degrees of freedom. This expansion yields an
infinite set of equations, where different pairs of Landau quantum
numbers $n^{\rm e}$, $n^{\rm p}$ are entangled, Eq.
(\ref{closed1}) being the equation for the $n^{\rm e}=n^{\rm p}=0$
component that decouples from this set in the limit $B=\infty$.

In the ultra-relativistic limit $P_0=P_3=0$ equation
(\ref{closed1}) is solved by the most symmetric Ansatz
$\Theta=I\Phi$, where $I$ is the unit matrix, and becomes
\begin{eqnarray}\label{2D}
\hspace{-0.3cm}\left(-\Box_2 + m^2\right)
\Phi(t,z)= 
{\rm i}16\pi\alpha D\left(t^2-z^2-\frac{P_\perp^2}{(eB)^2}\right)
\Phi(t,z).
\end{eqnarray}
 Here $\Box_2=
-\partial^2/\partial t^2+\partial^2/\partial z^2$ is the Laplace
operator. Equations for the rest two invariant coefficients, other
than the singlet $\Phi$, are considered in \cite{SU05}.

The ultraviolet singularity  on the light cone ($x^2=0$) of the
free photon propagator, $D(x^2)=-(\rm i/4\pi^2)1/x^2$, after one
substitutes this expression
 into eq. (\ref{2D}) taken for the lowest energy state
$P_\perp^2=0$, leads to falling to the center in the
Schr$\ddot{\rm o}$dinger-like
 differential equation\begin{eqnarray}\label{last3} -\frac{{\rm
d}^2\Psi(s)}{{\rm d} s^2}+\left(m^2-\frac
1{4s^2}\right)\Psi(s)=\frac{4\alpha}{\pi}\frac 1{s^2}\Psi(s),
\end{eqnarray}
to which the radial part of equation (\ref{2D}) is reduced in the
most
 symmetrical case, when the wave function $\Phi
(x)=s^{-1/2}\Psi(s)$ does not depend on the hyperbolic angle
$\phi$ in the spacelike region of the two-dimensional Minkowsky
space, $t=s\sinh\phi,\; z=s\cosh\phi,\;s=\sqrt{z^2-t^2}$.

The solution that decreases at $s\rightarrow \infty$ is given by
the McDonald function with imaginary index:
\begin{eqnarray}\label{mcdonald2}\Psi(s)=\sqrt{s}\;K_\nu(ms),\quad \nu=
{\rm i}\,2\sqrt{\alpha/\pi}\simeq 0.096\,{\rm i}\,.
\end{eqnarray} It behaves near the singular point $s=0$
as \begin{eqnarray}\label{behave} \left(\frac s{2}\right)^{1+
\nu}\frac 1{\Gamma(1+\nu)}-\left(\frac s{2}\right)^{1- \nu}\frac
1{\Gamma(1-\nu)}.\end{eqnarray} Here the Euler $\Gamma$-functions
appear. The falling to the center manifests itself \cite{QM} in
the complexity of the exponents in (\ref{behave}) that makes the
both asymptotic terms oscillating and equal in rights. The falling
to the center holds for any positive $\alpha$, the genuine value
$\alpha=1/137$ included, unlike the case without the magnetic
field, where nonphysically large values of the fine structure
constant, $\alpha>\pi/8$, are required to provide it
\cite{goldstein}, \cite{SU05}). Equation (\ref{last3}) is valid in
the interval
\begin{eqnarray}\label{interval} s_0\leq s\leq \infty,\quad s_0\gg
(eB)^{-1/2}.\end{eqnarray} Thus, the Larmour radius serves as a
regularizing length. According to \cite{shabad}, the singular
equation (\ref{last3}) should be considered as the generalized
eigenvalue problem with respect to $\alpha$. The operator in the
left-hand side is self-adjoint provided the boundary condition is
imposed, \begin{eqnarray}\label{stand} \Psi (s_0)=0 \,,
\end{eqnarray} that treats the Larmour radius as the lower
edge of the normalization box. The discrete eigenvalues
$\alpha_n(s_0)$ condense in the limit $B=\infty$ ($s_0=0$) to
become a continuum of states that make the Hilbert space of
vectors orthogonal with the singular measure $s^{-2}{\rm d}s$. The
latter fact allows us to normalize them to $\delta$-functions and
 interpret as free particles emitted and
absorbed by the singular center. In order that the Larmour radius
might be treated as the edge of the box it is necessary that $s_0$
be much smaller than the electron Compton length, $s_0\ll
m^{-1}\simeq 3.9\times 10^{-11}$ cm. Then the small-distance
asymptotic regime is reached, and the behavior of the system
"behind the horizon," $s<s_0,$ - where the two-dimensional
equations (\ref{closed1}), (\ref{2D}), and (\ref{last3}) are not
valid - is not important. In this way the existence of the limit
$s_0\rightarrow 0$, impossible in the standard theory, is achieved
\cite{footnote2}.

Beginning with a certain small value of the argument $ms$, the
function $K_\nu(ms)$ oscillates, as $s\rightarrow 0$, and takes
the zero value infinitely many times. To find the largest value of
$s_0$,
for which the boundary problem (\ref{last3}) and (\ref{stand}) can be
solved, one can use (\ref{behave}). Then eq. (\ref{stand}) reads
\begin{eqnarray}\label{spectre2}\nu\ln\frac {ms_0}{2}={\rm
i}\arg\Gamma(\nu+1)- {\rm i} \pi n,\quad n=0,\pm 1,\pm 2,...
\end{eqnarray} Since $|\nu|$ is small we may exploit the
approximation $\Gamma(1+\nu)\simeq 1-\nu C_{\rm E},$ where $C_{\rm
E}=0.577$ to get
\begin{eqnarray}\label{spectre3}\ln\left(\frac{ms_0}{2}\right)=-\frac
n{2} {\sqrt{\frac{\pi^3}{\alpha}}}-C_{\rm E}, \quad n=1,2...\quad
\end{eqnarray}
We have expelled the nonpositive integers $n$ from here, since
they would lead to the roots for $ms_0$ of the order of or larger
than unity in contradiction to the adopted condition $s_0\ll
m^{-1}$. For such values eq.(\ref{behave}) is not valid. It may be
checked that there are no other zeros of McDonald function, apart
from (\ref{spectre3}). The maximum value for $s_0$ is provided by
$n=1$:
\begin{eqnarray}\label{spectre4}
s_0^{\rm max}=\frac 2{m}\exp\left\{- \frac 1{2}
{\sqrt{\frac{\pi^3}{\alpha}}}-C_{\rm E}\right\} \simeq
10^{-14}m^{-1},
\end{eqnarray}
i.e., $s_0^{\rm max}$ is about 14 orders of magnitude
smaller than $m^{-1}$ and makes $\sim 10^{-25}$ cm. By demanding,
in accord with  (\ref{interval}), that the value of $s_0^{\rm
max}$ should exceed the Larmour radius
\begin{eqnarray}\label{B} s_0^{\rm max}\gg
L_B= (eB)^{-1/2}\quad {\rm or}\quad B\gg\frac 1{e\;(s_0^{\rm
max})^2}\end{eqnarray} one establishes, how large the magnetic
field should be in order that the boundary problem might have a
solution, in other words, that the point $P_0={\bf P}=0$ might
belong to the spectrum. Therefore, if the magnetic field exceeds
the maximum value of
\begin{eqnarray}\label{final} B_{\rm
max}\simeq \frac 1{ e(s_0^{\rm max})^2}\simeq
\frac{m^2}{4e}\exp\left\{\frac{\pi^{3/2}} {\sqrt{\alpha}}+2C_{\rm
E}\right\}\,
\end{eqnarray} the positronium
ground state with the center-of-mass 4-momentum equal to zero
exists \cite{footnote3}. The value of $B_{\rm max}$ is $1.6
\times 10^{28}\,B_0\sim 10^{42}$~G.  This is a few orders of
magnitude smaller than the magnetic field that may be in the
vicinity of superconductive cosmic strings \cite{W95}. Excited
positronium states may also reach the spectral point $P_\mu=0$,
but this occurs for magnetic fields, tens orders of magnitude
larger than (\ref{final}) - to be found in the same way from
(\ref{spectre3}) with $n=2,3...$

The ultrarelativistic state $P_\mu=0$ has the internal structure
of what was called a "confined state", belonging to kinematical
domain called "sector III" in \cite{shabad}, i.e., the one whose
wave function behaves as a standing wave combination of free
particles near the lower edge of the normalization box and
decreases as $\exp (-ms)$ at large distances. The effective "Bohr
radius", i.e., the value of $s$ that provides the maximum to the
wave function (\ref{mcdonald2}) makes $s_{\rm max}=0.17 m^{-1}$
This is  much less than the standard Bohr radius $(e^2m)^{-1}$.
The wave function is concentrated within the limits  $0.006
~m^{-1}<s<1.1~ m^{-1}$. But the effective region occupied by the
confined state is still much closer to $s=0$, since 
the probability density of the confined state is the wave function
squared $weighted ~with~ the~ measure$ $s^{-2}{\rm d} s$
$singular~ in~ the~ origin$ \cite{shabad} and is hence
concentrated near the edge of the normalization box $s_0\simeq
10^{-25}$cm.  The electric fields in the e$^+$e$^--$system at such
distances are about $10^{43}$ V/cm. There is no evidence that
the standard quantum theory (SQT) should be valid under such
conditions. This fact encourages the use of the theory of Ref.
\cite{shabad} above that differs from SQT in that it excludes the
short distances beyond the normalization box. (Note, nevertheless,
the reserves \cite{footnote1}, \cite{footnote2}.)

Compare the  value (\ref{final}) with the analogous value,
obtained earlier \cite{ShUs} by extrapolating the
semirelativistic result concerning the positronium binding energy
in a magnetic field to extreme relativistic
region:\begin{eqnarray}\label{old} \left.B_{\rm max}\right|_{\rm
nonrel}\simeq\frac{\alpha^2m^2}{e}\exp\left\{\frac{2\sqrt{2}}
{\alpha}\right\} \simeq 10^{164}\,B_0\,. \end{eqnarray} Such is
the magnetic field that makes the binding energy of the lowest
energy state equal to $-2m$. We see that the relativistically
enhanced attraction has resulted in a drastically lower value of
the maximum magnetic field.

Let us estimate possible effects of radiative corrections (see
Ref. \cite{SUMarch} for details).

May the vacuum polarization screen the attraction force between
the electron and positron in such a way as to prevent the
positronium collapse? The three photon polarization eigenmodes
give the contributions into the photon propagator
\cite{batalin},\cite{kniga}
 to be used in the BS equation,
\begin{eqnarray}\label{mode} D_{ a}(x)=-\frac 1{(2\pi)^4}\int\frac
{\exp ({\rm i}kx){\rm d}^4k}{k^2+\kappa_{a} (k)},\quad a=1,2,3,
\end{eqnarray}which contain  the polarization
tensor eigenvalues $\kappa_a(k)$ in the denominator. When
calculated \cite{batalin} in the one-loop approximation,
$\kappa_{1,3}$ grow with the field as $(\alpha/3\pi)\ln (B/B_0)$,
but this remains yet small ($\sim 0.04$) for the fields of the
order of (\ref{final}). The interaction between charged particles
carried by  these two photon modes remains practically the same as
it was without the vacuum polarization and thus continues to
support the collapse. However,  the  eigenvalue  $\kappa_{2}$,
besides the logarithmic growth, also contains a  term that
increases with $B$ linearly \cite{kr}, \cite{kniga}:
\begin{eqnarray}
\kappa_2(k) =\frac{\alpha Bm^2(k_0^2-k_3^2)}{\pi B_0}\exp \left(-
\frac{k_\bot^2}{2m^2}\frac{B_0}{B} \right)\nonumber\\
\times\int_{-1}^1\frac{(1-\eta^2)\rm d
\eta}{4m^2-(k_0^2-k_3^2)(1-\eta^2)}\label{2}\end{eqnarray} and
might seemingly screen the interaction carried by the mode 3
photon. Here $k_0$ is the photon energy, $k_3$ and ${\bf k}_\perp$
are its momentum components along and transverse the magnetic
field, $k^2=k_0^2-{\bf k}^2$. However, once the electron-positron
separation in the plane perpendicular to the magnetic field ${\bf
x}_\perp ={\bf x}_\perp^{\rm e}-{\bf x}_\perp^{\rm p}$ is
restricted to the domain inside the Larmour radius
$|x_\perp|\ll(eB)^{-1/2}$, the integration in (\ref{mode}) gets
essential contribution from  large $|k_\perp|\gg (eB)^{1/2}$.  In
this domain the exponential in (\ref{2}) suppresses the linearly
growing term in the denominator of $D_2$. This is how the
exponentially strong spatial dispersion  opposes the screening
\cite{SUMarch}.

Among the mass radiative corrections, the so-called ln$^2$ terms
are potentially dangerous for the present gap-shrinking effect,
since they yield a competing growth of the corrected electron
mass, essential at the scale of (\ref{final}). Substituting the
one-loop corrected \cite{jancovici} electron mass $\widetilde{m}$,
\begin{eqnarray}\label{mass}\widetilde{m}\simeq
m\left(1+\frac\alpha{4\pi}\ln^2\frac{B}{B_0}\right),\end{eqnarray}
for $m$, and $L_B=(eB)^{-1/2}$ for $s_0^{\rm max}$ into equation
(\ref{spectre4}), we estimate that the maximum field increases
only by a factor of $\sim 10$. The most recent results concerning
the mass correction that take into account the vacuum polarization
diagrams inside the mass operator \cite{osipov} read that the
ln$^2$ terms do not, as a matter of fact, appear at all, 
thanks to the presence of the growing term (\ref{2}).

At $B=B_{\rm max}$ the total energy and momentum of a positronium
in the ground state are zero. This state is not separated from the
vacuum by an energy gap, and it is the one of maximum symmetry in
the coordinate and spin space. Hence, it may be related to the
vacuum and describe its structure.

What happens when the magnetic field exceeds the maximum
value (\ref{final})? To answer this question  one would have to
solve (a more complicated) BS equation with $P_{\parallel}$
nonzero and spacelike. 
In that case we transit from the "sector III of
confined states" 
to the
"deconfinement sector IV" \cite{shabad} where solutions are free
waves both near $s=0$ and $s=\infty$ and correspond to delocalized
states of mutually free electron and positron - each on its
Larmour orbit - moving along the magnetic field. These are capable
of screening the magnetic field and put a limit to its further
growing. A more detailed discussion of this hypothesis that traces
an analogy with the known problem of a Dirac electron in the
Coulomb field of a supercharged nucleus \cite{popov} and also
dwells on the structure of the corresponding
translationally-non-invariant vacuum state is presented in
\cite{SU05}. For the present, we state that the hypervalue
(\ref{final}) is such a value of the magnetic field, the exceeding
of which would cause restructuring of the vacuum and demand a
revision of QED.

The authors are grateful to V. Gusynin, V. Miransky and N. Mikheev
for valuable
discussions. This work was supported by the Russian
 Foundation for Basic Research
 (project no 05-02-17217) and the President of Russia Programme
 (LSS-4401.2006.2), as well as
by the Israel Science Foundation of the Israel Academy of Sciences
and Humanities.

\vfill\eject

\begin{thebibliography}{20}
\bibitem{footnote1}{Still,
the final result (\ref{final}) may be obtained keeping to the
standard cut-off philosophy.}
\bibitem{footnote2} Alternatively, we might handle $s_0$
as the cut-off parameter, like in \cite{QM}, by replacing the
singularity $s^{-2}$ by a a constant $s_0^{-2}$and applying the
zero boundary condition in $s_0=0$. The result (\ref{final}) below
is but a little sensitive
 to this alteration \cite{SU05}.
\end{thebibliography}
\end{document}